\documentclass[prd,preprint,nofootinbib]{revtex4} 
\usepackage{graphicx,amsmath,amsfonts,amssymb,revsymb,dcolumn} 
\usepackage[utf8]{inputenc}
\usepackage[toc,page]{appendix}

\newcommand{\be}{\begin{equation}}
\newcommand{\ee}{\end{equation}}
\newcommand{\bea}{\begin{eqnarray}}
\newcommand{\eea}{\end{eqnarray}}

\begin{document}

\title{On the consistency of the Lagrange multiplier method in classical mechanics}

\author{Nivaldo A.  \surname{Lemos}} \email{nivaldolemos@id.uff.br} \author{Marco \surname{Moriconi}} \email{mmoriconi@id.uff.br}
\affiliation{Instituto de Física, Universidade Federal Fluminense,
Campus da Praia Vermelha, Niterói, 24210-340, RJ, Brazil.}

\date{\today}


\date{\today}

\begin{abstract}
\noindent Problems involving rolling without slipping or no sideways skidding, to name a few, introduce velocity-dependent constraints that can be efficiently treated by the method of Lagrange multipliers in the Lagrangian formulation of the classical equations of motion. In doing so one finds, as a bonus, the constraint forces, which must be independent of the solution of the equations of motion, and can only depend on the generalized coordinates and velocities, as well as time. In this paper we establish conditions the Lagrangian and the constraints should obey in order to guarantee that the constraint forces can be obtained consistently.
\end{abstract}

\maketitle

\section{Introduction}

Sometimes the mathematical treatment of  certain  physical problems seems so well established, so ingrained in the tradition of theoretical physics  and  vindicated  by so many examples, that it is widely believed to provide physically correct answers in all instances. But even if a physical theory  appears to be definitely settled,  a closer examination of its mathematical foundations is not  a waste of time because it may reveal  overlooked conditions  for the theory to be free from physical inconsistencies.  A case in point is the application of the method of Lagrange multipliers in classical mechanics  to deal with a large class of constrained systems.  

In classical mechanics one often encounters problems involving   rolling without slipping, such as that of a rolling coin on a slanting table  \cite{Lemos} or a ball on a rough surface of revolution \cite{English,White}, and also those entailing  the requirement  of not skidding sideways, as in the case of  a skate or a bicycle \cite{Lemos,Neimark}, a two-wheeled cart driven by electrostatic forces \cite{Gatland} or  a snakeboard  \cite{Janova}. The  restrictions that characterize the allowed motions of these systems are velocity-dependent constraints, making them a class of constraints worth our attention. The equations that express these constraints are first-order ordinary differential equations that depend linearly on the velocities associated with the coordinates chosen to describe the configurations of the system. In most cases these constraints are nonholonomic, and they are usually dealt with by the technique of Lagrange multipliers. A bonus of this approach is that the constraint forces are also found in the process of solving the equations of motion.  Even holonomic constraints can be advantageously tackled by this method when one is interested in determining the constraint forces.

Forces are functions only of positions, velocities and time; they cannot depend on accelerations or even higher derivatives of the positions \cite{Zangwill}. This imposes a physical consistency requirement on the method of Lagrange multipliers that, as far as we know,  is not discussed  in the standard textbooks. Here we state clearly what this consistency requirement is and find  conditions under which the technique of Lagrange multipliers is physically consistent. 

\section{The Lagrange multiplier method in classical mechanics}\label{Lagrangemultipliermethod}

Consider an $n$-degree-of-freedom mechanical system described by coordinates $q_1, \ldots , q_n$ subject to a set of $p<n$ mutually independent differential constraints that depend linearly on the velocities, namely
\begin{equation}
\label{vinculos-lineares-velocidades-diferenciais}
\sum_{k=1}^n\, a_{lk} dq_k  +  \alpha_{l}dt  = 0\, , \,\,\,\,\,\,\,\,\,\,
l=1,\ldots ,p\, .
\end{equation}
The coefficients $a_{lk}$ and $\alpha_{l}$ do not depend on the velocities; that is, they are functions of $q_1, \ldots , q_n, t$ alone: $a_{lk} = a_{lk}(q,t), \, \alpha_{l} = \alpha_{l}(q,t)$.
In order to ensure the mutual independence of the above constraint equations  we assume that the $p \times n$ rectangular matrix 
\begin{equation}
\label{matrix-A-constraints}
\boldsymbol{\mathsf{A}} = 
                   \left( \begin{array}{ccccc}
                  a_{11} & a_{12}  &  a_{13} & \cdots & a_{1n}   \\                         
										\vdots & \vdots & \vdots & \vdots & \vdots\\
										a_{p1} & a_{p2} & a_{p3} & \cdots & a_{pn}                 
                  \end{array}
                  \right)  
\end{equation}
has rank $p$ for all $q,t$. This means that, with a suitable numbering of the coordinates,
\begin{equation}
\label{det-submatrix-A-constraints} 
                \det   \left( \begin{array}{ccc}
                  a_{11} &  \cdots & a_{1p}   \\                         
										\vdots & \vdots & \vdots \\
										a_{p1} & \cdots & a_{pp}                 
                  \end{array}
                  \right)  \neq 0 \,\,\,\,\, \mbox{for all} \,\,\,\,\, q,t.
\end{equation}
By way of illustration, for an upright coin rolling without slipping on a horizontal plane the constraints are \cite{Lemos} 
\begin{equation}
\label{constraints-coin}
{\dot x} - R {\dot \phi}\cos \theta = 0  , \,\,\,\,\,\,\,\,\,\,
{\dot y} - R {\dot \phi}\sin \, \theta = 0 . 
\end{equation}
With $q_1=x, q_2=y, q_3=\phi, q_4=\theta$ and upon multiplication of these constraint  equations by $dt$,  
 it follows that
\begin{equation}
\label{matrix-A-constraints-coin}
\boldsymbol{\mathsf{A}} = 
                   \left( \begin{array}{cccc}
                    1 &\,\,\,\, 0 \,\,\,\, & - R \cos \theta & \,\,\,\, 0   \\                         
										0 & \,\,\,\, 1 \,\,\,\,  & - R \sin \theta & \,\,\,\, 0                  
                  \end{array}
                  \right)  .
\end{equation}
This is a coordinate-dependent matrix whose rank is $2$ for all $q,t$.

Equation (\ref{det-submatrix-A-constraints}), combined with the fact  that transposition does not change the determinant of a matrix,  has the following important consequence: if  $\boldsymbol{\mathsf{s}}$ is a $p$-dimensional column vector and $\boldsymbol{\mathsf{A}}^T \boldsymbol{\mathsf{s}} =0$ then $ \boldsymbol{\mathsf{s}} =0$. This fact will be used in the proof of our main result in Section \ref{consistency-proof}.

Let $L(q, {\dot q},t)$ be the Lagrangian for the system, {\it written as if there were no constraints}. The Lagrange multiplier method to take into account the constraints (\ref{vinculos-lineares-velocidades-diferenciais}) in Hamilton's principle \cite{Lemos} starts by noting that the variations $\delta q_k$ that enter Hamilton's variational principle   are virtual displacements that, because of
(\ref{vinculos-lineares-velocidades-diferenciais}), obey
\begin{equation}
\label{vinculos-lineares-velocidades-virtual}
\sum_{k=1}^n\, a_{lk} \delta q_k   = 0\, , \,\,\,\,\,\,\,\,\,\,
l=1,\ldots ,p\, 
\end{equation}
inasmuch as $dt =0$ for virtual displacements.  Equation (\ref{det-submatrix-A-constraints}) implies that the first $p$ variations  $\delta q_1, \dots , \delta q_p$ can be expressed in terms of the remaining ones $\delta q_{p+1}, \dots , \delta q_n$, which are independent and arbitrary.  The introduction of Lagrange multipliers  to take the subsidiary conditions (\ref{vinculos-lineares-velocidades-virtual}) into account in Hamilton's principle 
leads to the equations of motion \cite{Lemos}  
\begin{equation}
\label{equacoes-Lagrange-com-multiplicadores-de-Lagrange}
\frac{d}{dt}
\Bigl(\frac{\partial L}{\partial  {\dot q_k}} \Bigr) -
\frac{\partial L}{\partial q_k} =
\sum_{l=1}^p\,{\lambda}_l a_{lk}
\, , \,\,\,\,\,\,\,\,\,\,
k=1,\ldots ,n\, ,
\end{equation}
where $\lambda_1, \ldots , \lambda_p$ are  Lagrange multipliers.

Of course, the differential constraints (\ref{vinculos-lineares-velocidades-diferenciais}) can be written in the equivalent form of a set of $p$ first-order ordinary differential equations:  
\begin{equation}
\label{vinculos-lineares-velocidades-equacoes-diferenciais}
\sum_{k=1}^n\, a_{lk} {\dot q}_k  +  \alpha_l  = 0\, , \,\,\,\,\,\,\,\,\,\,
l=1,\ldots ,p\, .
\end{equation}

Equations  (\ref{equacoes-Lagrange-com-multiplicadores-de-Lagrange}) and (\ref{vinculos-lineares-velocidades-equacoes-diferenciais}) comprise a set  of $n+p$
equations for $n+p$ unknowns --- the $n$ coordinates $q_1, \ldots , q_n$ and the $p$ Lagrange multipliers $\lambda_1, \ldots , \lambda_p$ --- which uniquely determine the motion of the system  once initial conditions obeying the constraints  are stipulated. In terms of the Lagrange multipliers, which are determined in the process of solving equations  (\ref{equacoes-Lagrange-com-multiplicadores-de-Lagrange}) and (\ref{vinculos-lineares-velocidades-equacoes-diferenciais}),  the generalized constraint forces $Q^{\prime}_k$  responsible for the enforcement of (\ref{vinculos-lineares-velocidades-equacoes-diferenciais}) are given by \cite{Lemos}
\begin{equation}
\label{forcas-de-vinculo-em-termos-multiplicadores-Lagrange}
Q^{\prime}_k= \sum_{l=1}^p\,{\lambda}_l a_{lk} \, , \,\,\,\,\,\,\,\,\,\,
k=1,\ldots ,n\,  .
\end{equation}

The technique of Lagrange multipliers can be extended to nonholonomic  constraints of the more general  form
\begin{equation}
\label{non-holon-more-general}
g_l(q, {\dot q}, t) =0 \, , \,\,\,\,\,\,\,\,\,\,
l=1,\ldots ,p\, .
\end{equation}
In this case    
the accepted equations of motion are \cite{Rund,Saletan}
\begin{equation}
\label{equacoes-Lagrange-com-multiplicadores-de-Lagrange-general}
\frac{d}{dt}
\Bigl(\frac{\partial L}{\partial  {\dot q_k}} \Bigr) -
\frac{\partial L}{\partial q_k} =
\sum_{l=1}^p\,{\lambda}_l \frac{\partial g_l}{\partial {\dot q}_k}
\, , \,\,\,\,\,\,\,\,\,\,
k=1,\ldots ,n\, .
\end{equation}
These equations of motion define the d'Alembert-Lagrange theory of nonholonomic systems.  It is clear that equations (\ref{equacoes-Lagrange-com-multiplicadores-de-Lagrange-general})  reduce to the previous equations of motion (\ref{equacoes-Lagrange-com-multiplicadores-de-Lagrange}) when the functions $g_l$ depend linearly on the velocities.  All of our forthcoming  results remain valid for this more general case by just putting $a_{lk} = \partial g_l/ \partial {\dot q}_k$,   despite the fact that the $a_{lk}$ will depend on the velocities, as long as it is assumed that the matrix $\boldsymbol{\mathsf{A}} =(\partial g_l/ \partial {\dot q}_k)$ has rank $p$ for all $q, {\dot q},t$. 

It is also worth pointing out that the {\it physical} Lagrange multiplier method to deal with nonholonomic constraints  in classical mechanics \cite{Rund,Saletan,Flannery1,Flannery2,Flannery3} leads to equations of motion that are different from  those derived in the treatment of the corresponding {\it mathematical} problem of Lagrange  in the calculus of variations \cite{Bliss}. The latter gives rise to the so-called vakonomic mechanics \cite{Ray,Goedecke,Kozlov,Borisov},  which features ${\dot \lambda_l}$   in the equations of motion in addition to the Lagrange multipliers $\lambda_l$ themselves. The reader is referred to the Appendix for a short overview of this controversial topic.

 \section{A consistency requirement}\label{Consistency-Requirement}

In Newtonian mechanics the most general forces are functions of coordinates, velocities and time, and this must be true of  the constraint forces (\ref{forcas-de-vinculo-em-termos-multiplicadores-Lagrange}). Therefore, {\it without having to explicitly solve for the motion}, it should  be possible to express the Lagrange multipliers solely in terms of $q,{\dot q},t$. The accelerations cannot appear
in the solution for the Lagrange multipliers.

 Particular examples \cite{Lemos,Neimark} show that by combining the equations of motion (\ref{equacoes-Lagrange-com-multiplicadores-de-Lagrange}) with the constraint equations (\ref{vinculos-lineares-velocidades-equacoes-diferenciais}) one  manages to express $\lambda_1, \ldots , \lambda_p$ in terms of   $q,{\dot q},t$ only.
Thus, for consistency, one must prove that this can be achieved in general.
To this end, let us start by writing the equations of motion in the explicit form 
\begin{equation}
\label{eqs-motion-explicit}
 \sum_{j=1}^n \frac{\partial^2 L}{\partial{\dot q}_j\partial{\dot q}_k}\,{\ddot q}_j +
\sum_{j=1}^n \frac{\partial^2 L}{\partial q_j\partial {\dot q}_k}\,{\dot q}_j
+ \frac{\partial^2 L}{\partial t \partial{\dot q}_k} - 
\frac{\partial L}{\partial q_k} = \sum_{l=1}^p\,{\lambda}_l a_{lk}\, , \,\,\,\,\,\,\, k=1,\ldots , n
\, . 
\end{equation}

We assume that we are dealing with a regular Lagrangian, which means that the  Hessian matrix
 $\boldsymbol{\mathsf{W}}=(W_{jk})$ with  elements
\begin{equation}
\label{elementos-matriz-hessiana}
W_{kj} = W_{jk}=
\frac{\partial^2 L}{\partial {\dot q}_j\partial {\dot q}_k}
\,  
\end{equation}
is nonsingular:
\begin{equation}
\label{determinante-matriz-hessiana-diferente-zero}
\det  \boldsymbol{\mathsf{W}} \neq 0  \,\,\,\,\,\,\,\,\,\, \mbox{for all} \,\,\,\,\,\,\,\,\,\, (q, \dot q, t) \in {\cal D},
 \end{equation}
where ${\cal D}$ is the Lagrangian's domain of definition. This condition is sufficient to allow the usual Lagrange's equations to be solved for the accelerations   and put in the standard form  ${\ddot q}_k = f_k(
q,{\dot q},t)$. Then, if (\ref{determinante-matriz-hessiana-diferente-zero}) holds and the Lagrangian is smooth, the existence and uniqueness theorems of the theory of ordinary differential equations apply  and establish  that  once we fix the coordinates  and velocities at some initial instant Lagrange's equations uniquely determine the motion  at least locally in time, a physical requisite one cannot do without. It is also worth noting that  condition (\ref{determinante-matriz-hessiana-diferente-zero})  is all that is required in order for the Hamiltonian formalism associated with the Lagrangian $L$ to be locally well defined \cite{Lemos}. To ensure the global existence of the Hamiltonian formalism a stronger condition is needed:  the Lagrangian should be hyperregular \cite{Marsden}.

Equations (\ref{eqs-motion-explicit})  can be conveniently recast as 
\begin{equation}
\label{eqs-motion-minus-lambda}
 \sum_{l=1}^p\,(-{\lambda}_l) a_{lk} + \sum_{j=1}^n W_{kj}{\ddot q}_j  = b_k\, , \,\,\,\,\,\,\, k=1,\ldots , n
\, ,
\end{equation}
where $b_k = b_k(q, {\dot q}, t)$.

Differentiating the constraint equations (\ref{vinculos-lineares-velocidades-equacoes-diferenciais}) with respect to time we arrive at
\begin{equation}
\label{vinculos-accelerations}
\sum_{k=1}^n\, a_{lk} {\ddot q}_k   = c_l \, , \,\,\,\,\,\,\,\,\,\,
l=1,\ldots ,p\, ,
\end{equation}
where $c_l = c_l (q, {\dot q}, t)$. In the following, all that matters is the fact that $b_k$ and $c_l$ depend only on coordinates and velocities, not their specific functional form. 

Let $\boldsymbol{\mathsf{B}}$ be the $(n+p) \times (n+p)$ matrix defined in terms of the matrices $\boldsymbol{\mathsf{A}}$ and $\boldsymbol{\mathsf{W}}$ by
\begin{equation}
\label{matriz-B}
 \boldsymbol{\mathsf{B}}   =
                   \left( \begin{array}{lc}
                  \boldsymbol{\mathsf{O}} &   \boldsymbol{\mathsf{A}}   \\                         
											 \boldsymbol{\mathsf{A}}^T &   \boldsymbol{\mathsf{W}} \\                 
                  \end{array}
                  \right)  
\end{equation}
where $\boldsymbol{\mathsf{O}}$ is the $p \times p$ zero matrix. 
Let us also introduce the $(n+p)$-dimensional column vectors
\begin{equation}
\label{column-vectors}
 \boldsymbol{\mathsf{v}}   =
                   \left( \begin{array}{c}
                              -\lambda_1   \\           												    
															\vdots  \\
															-\lambda_p \\
															{\ddot q}_1 \\
															 \vdots  \\
															{\ddot q}_n
																										
                  \end{array}
                  \right)  \, , \,\,\,\,\,\,\,\,\,\,\,\,\,\,\,\,\,\,\,\,\,\,\,\,\,\,\,\,\,\, \boldsymbol{\mathsf{e}}   =
                   \left( \begin{array}{c}
                              c_1   \\           												    
															\vdots  \\
															c_p \\
															b_1 \\
															 \vdots  \\
															b_n
																										
                  \end{array}
                  \right) \, .
\end{equation}
In terms of these column vectors and the matrix (\ref{matriz-B}), the set of $n+p$ equations  (\ref{eqs-motion-minus-lambda}) and (\ref{vinculos-accelerations})  can be summarized as
\begin{equation}
\label{n+p-equations-matrix-form}
\boldsymbol{\mathsf{B}} \boldsymbol{\mathsf{v}} = \boldsymbol{\mathsf{e}}\, .
\end{equation}
Note that both the matrix  $\boldsymbol{\mathsf{B}}$ and the vector $\boldsymbol{\mathsf{e}}$ depend only on $q, {\dot q}, t$. Therefore, if $\boldsymbol{\mathsf{B}}$ is a nonsingular matrix, the unique solution of equation (\ref{n+p-equations-matrix-form}) for $\boldsymbol{\mathsf{v}}$ is $\boldsymbol{\mathsf{v}} = \boldsymbol{\mathsf{B}}^{-1} \boldsymbol{\mathsf{e}}$, which gives  $\lambda_1, \dots ,\lambda_p$ as functions of $q, {\dot q}, t$ alone. Incidentally, the accelerations are also determined in the form ${\ddot q}_k = f_k (q, {\dot q},t)$, which, if  the Lagrangian as  well as the constraint coefficients $a_{lk}$ and $\alpha_l$ are smooth functions,  guarantees existence and uniqueness for the motion at least locally in time  once initial coordinates and velocities
 obeying the constraints are specified.

It follows that in order for the method of Lagrange multipliers in classical mechanics to yield physically meaningful results one must demand that
\begin{equation}
\label{consistency-requirement}
\det \boldsymbol{\mathsf{B}} \neq 0 \,\,\,\,\,\,\,\,\,\, \mbox{for all} \,\,\,\,\,\,\,\,\,\, (q, {\dot q}, t) \in {\cal D}
\end{equation}
as an essential consistency requirement. To be precise, the condition $\det \boldsymbol{\mathsf{B}} \neq 0$ ensures that the Lagrange multipliers are physically sensible functions of $q, \dot{q}, t$.  

Note that Eq.(\ref{consistency-requirement}) is only a sufficient condition: if it fails
it may still be possible to determine the Lagrange multipliers in terms of $q, {\dot q},t$ only. 
But even if this is the case, 
$\det \boldsymbol{\mathsf{B}} = 0$ implies that the $n+p$ equations (\ref{n+p-equations-matrix-form}) for the $\lambda$s and ${\ddot q}$s are not  mutually independent. This leads to
 the appearence of at least one equation relating only coordinates and velocities,  meaning that some initial conditions cannot be satisfied by the general solution to the equations of motion. 
In order to justify this assertion, let us start by noting that the matrix $\boldsymbol{\mathsf{B}} $ defined by equation (\ref{matriz-B}) is symmetric. Suppose $\det \boldsymbol{\mathsf{B}} = 0$ for all values of $q, {\dot q},t$ in some set $\cal O$. Then there exists at least one $(n+p)$-dimensional nonzero  vector $\boldsymbol{\mathsf{u}}(q, {\dot q},t)$ such that $ \boldsymbol{\mathsf{B}}\boldsymbol{\mathsf{u}}=0$ on $\cal O$,  which is equivalent to
\begin{equation}
\label{B-zero-eigenvalue-X}
\boldsymbol{\mathsf{u}}^T\boldsymbol{\mathsf{B}} =0
 \,\,\,\,\,\,\,\,\,\, \mbox{for all} \,\,\,\,\,\,\,\,\,\, (q, {\dot q},t) \in  {\cal O},
\end{equation}
where the symmetry of $\boldsymbol{\mathsf{B}}$ has been used.
Combining the above equation with  $\boldsymbol{\mathsf{B}}\boldsymbol{\mathsf{v}} = \boldsymbol{\mathsf{e}}$ we are led to
\begin{equation} 
\label{Bv=e-in-components}
\boldsymbol{\mathsf{u}}^T \boldsymbol{\mathsf{e}} = \boldsymbol{\mathsf{u}}^T\boldsymbol{\mathsf{B}}\boldsymbol{\mathsf{v}} = 0 \,\,\,\,\,\,\,\,\,\, \mbox{for all} \,\,\,\,\,\,\,\,\,\, (q, {\dot q},t) \in  {\cal O}.
\end{equation}
In terms of components, the condition $\boldsymbol{\mathsf{u}}^T \boldsymbol{\mathsf{e}} = 0$ on $\cal O$ reads
\begin{equation}
\label{additional-constraint}
\sum_{r=1}^{n+p}u_r(q, {\dot q},t)e_r(q, {\dot q},t) =  0 \,\,\,\,\,\,\,\,\,\, \mbox{for all} \,\,\,\,\,\,\,\,\,\, (q, {\dot q},t) \in  {\cal O}.
\end{equation}
Unless it is identically satisfied owing to an accident,  this equation provides an additional constraint.
A consequence of this additional constraint is that some allowed initial conditions cannot be fulfilled, which is a physical inconsistency.
The ocurrence of this state of affairs will be shown explicitly by means of Example 1 in Section \ref{status-positivity}.  

In short, even though it is possible to have physically consistent Lagrange multipliers without (\ref{consistency-requirement}) being satisfied, failure to satisfy (\ref{consistency-requirement}) is almost certain to bring about some inconsistency.


The Lagrangians associated with most mechanical systems are not only regular but also  smooth functions of $q,{\dot q}, t$. This smoothness allied to the regularity condition (\ref{determinante-matriz-hessiana-diferente-zero})  is sufficient 
to make sure that  Lagrange's equations in their usual form locally yield a unique motion under standard initial conditions. It is also enough for the  derivation of a  Hamiltonian and Hamilton's equations from the given Lagrangian $L$, albeit only locally. Therefore, it would be quite natural and reasonable to  expect that it is also sufficient for the technique of Lagrange multipliers to lead to physically sensible results, that is, for the validity of (\ref{consistency-requirement}). In the following we show that this is not the case.   

\section{A consistency proof}\label{consistency-proof}

The standard Lagrangian in classical mechanics is of the form $L = T - V$,  where $T$ is a positive quadratic form in the variables ${\dot q}_1, \ldots , {\dot q}_n$ and $V$ does not depend on the velocities. This implies that the associated Hessian matrix $ \boldsymbol{\mathsf{W}}$ is not only real, symmetric and nonsingular, but also positive, since, on physical grounds, the kinetic energy is positive for any set of velocities, that is,
\begin{equation}
\label{W-positive}
\boldsymbol{\mathsf{t}}^T \boldsymbol{\mathsf{W}} \boldsymbol{\mathsf{t}} > 0 \,\,\,\,\,\,\, \mbox{if} \,\,\,\,\, \boldsymbol{\mathsf{t}} \neq 0 \,\,\,\,\,\,\,\,\,\, \mbox{and} \,\,\,\,\,\,\,\,\,\, \boldsymbol{\mathsf{t}}^T \boldsymbol{\mathsf{W}} \boldsymbol{\mathsf{t}} =0  \,\, \Longleftrightarrow \,\,  \boldsymbol{\mathsf{t}} =0,
\end{equation}
where $\boldsymbol{\mathsf{t}}$ is an $n$-dimensional column vector. It is to be noted that a real symmetric positive matrix is automatically nonsingular, since its eigenvalues are all real and greater than zero. Moreover, its inverse is also positive, since the eigenvalues of the inverse matrix are just the reciprocals of the eigenvalues of the original positive matrix.

The previous  remarks, and the significant fact that the method of Lagrange multipliers is successful in all known physical examples,  lead us to suspect that one cannot prove in general that the matrix (\ref{matriz-B}) is nonsingular unless the Hessian matrix is assumed to be positive.

\vspace{.5cm}

\noindent {\bf Proposition} \hspace{.5cm} Let $p$ and $n$ be positive integers such that $p < n$. If the $p \times n$ matrix $\boldsymbol{\mathsf{A}}$ has rank $p$ and  the $n \times n$ matrix $\boldsymbol{\mathsf{W}}$ is positive then the $(n+p) \times (n+p)$ matrix $\boldsymbol{\mathsf{B}}$ defined by equation (\ref{matriz-B}) is invertible.

\noindent {\bf Proof} \hspace{.5cm} In order to prove that $\boldsymbol{\mathsf{B}}$ is invertible we must show that $\det{\boldsymbol{\mathsf{B}} \neq 0}$, that is, $\boldsymbol{\mathsf{B}}$ has no zero eigenvalue. We show this by proving that  $\boldsymbol{\mathsf{B}}  \boldsymbol{\mathsf{v}}= 0$ implies $\boldsymbol{\mathsf{v}} =0$.

Let us write $\boldsymbol{\mathsf{v}}$ as
\begin{equation}
\label{column-vectors2}
 \boldsymbol{\mathsf{v}}   =
                   \left( \begin{array}{c}
                              \boldsymbol{\mathsf{s}}   \\
                              \boldsymbol{\mathsf{t}}
                  \end{array}
                  \right)
\end{equation}
where $\boldsymbol{\mathsf{s}}$ and $\boldsymbol{\mathsf{t}}$ are $p$-dimensional and $n$-dimensional vectors, respectively.
The equation $\boldsymbol{\mathsf{B}} \boldsymbol{\mathsf{v}} = 0$ becomes
\begin{eqnarray}
    \boldsymbol{\mathsf{A}} \boldsymbol{\mathsf{t}} = 0  ; \label{cond1} \\
    \boldsymbol{\mathsf{A}}^T \boldsymbol{\mathsf{s}} + \boldsymbol{\mathsf{W}} \boldsymbol{\mathsf{t}} = 0 \label{cond2}   .
\end{eqnarray}
Since $\boldsymbol{\mathsf{W}}$ is invertible, applying $\boldsymbol{\mathsf{s}}^T \boldsymbol{\mathsf{A}}\boldsymbol{\mathsf{W}}^{-1}$ on (\ref{cond2}) and using (\ref{cond1}), we obtain
\begin{equation}
    \boldsymbol{\mathsf{s}}^T  \boldsymbol{\mathsf{A}}\boldsymbol{\mathsf{W}}^{-1} \boldsymbol{\mathsf{A}}^T\boldsymbol{\mathsf{s}} = 0 , \label{sawas}
\end{equation}
which, together with the positivity of $\boldsymbol{\mathsf{W}}^{-1}$, implies $\boldsymbol{\mathsf{A}}^T \boldsymbol{\mathsf{s}} = 0$. As pointed out in Section \ref{Lagrangemultipliermethod}, since $\boldsymbol{\mathsf{A}}$ has rank $p$, this implies $\boldsymbol{\mathsf{s}}=0$ which, together with (\ref{cond2}), implies $\boldsymbol{\mathsf{t}} = 0$.\hfill $\square$

\vspace{.5cm}

Thus, if the conditions stated in  this proposition are fulfilled, the Lagrange multipliers are uniquely determined as functions only of $q, {\dot q}, t$ and so are the constraint forces   (\ref{forcas-de-vinculo-em-termos-multiplicadores-Lagrange}).

The proof of the above proposition shows only that the positivity  hypothesis  on $\boldsymbol{\mathsf{W}}$ together with the full rank of the constraint matrix  $\boldsymbol{\mathsf{A}}$  guarantees that the Lagrange multipliers can be expressed as functions of $q, {\dot q}, t$ and nothing else. But is the positivity of $\boldsymbol{\mathsf{W}}$ a condition one can go without? It turns out that if the  positivity   of $\boldsymbol{\mathsf{W}}$  is relinquished we run the risk of getting in trouble, as discussed in the next section.

\section{Status of the positivity of the Hessian matrix}\label{status-positivity}	

In  the typical  examples of physical interest the Hessian matrix $\boldsymbol{\mathsf{W}}$ is positive. Let us investigate what can happen when $\boldsymbol{\mathsf{W}}$ is not positive by means of the study of two specific cases.    Example 1 illustrates the case alluded to before in which $\det \boldsymbol{\mathsf{B}} = 0$, but the multipliers can still be solved for in terms of coordinates, velocities and time (but other inconsistencies arise).  
Example 2 illustrates a case in which the Hessian matrix is non-positive but physically consistent results for the Lagrange multipliers can still be obtained.  

Our first example shows that, as a rule,  the positivity hypothesis on  
$\boldsymbol{\mathsf{W}}$ cannot be dispensed with.

{\bf Example 1.}  Let the Lagrangian for a three-degree-of-freedom system be 
\begin{equation}
\label{lagrangian-counterexample}
L = \frac{{\dot q}_1^2}{2} + \frac{{\dot q}_2^2}{2} + {\dot q}_1{\dot q}_2 + {\dot q}_2{\dot q}_3 - \frac{1}{2} (q_1^2 + q_2^2  + q_3^2 ) \, .
\end{equation} 
For the corresponding Hessian matrix we have
\begin{equation}
\label{Hessian-counterexample}
 \boldsymbol{\mathsf{W}}   =
                   \left( \begin{array}{ccc}
                     1 & 1 & 0    \\                         
										 1 & 1 & 1  \\ 
										 0 & 1 & 0 							
                  \end{array}
                  \right)  \, .
\end{equation}
This matrix is invertible since its determinant equals $-1$, which means that the Lagrangian (\ref{lagrangian-counterexample}) is regular. However,  it is not positive. Indeed, since $\boldsymbol{\mathsf{W}}$ is real and symmetric, its eigenvalues are all real, and they cannot be all positive, for their product is $-1$. 


Let us assume that the system is subject to the two  constraints
\begin{equation}
\label{constraints-counterexample}
{\dot q}_1 + q_3 = 0 , \,\,\,\,\,\,\,\,\,\,\,\,\,\,\,\,\,\,\,\, {\dot q}_2 - q_3 = 0 ,
\end{equation} 
which  can be readily shown to be nonholonomic  by an appeal to the Frobenius theorem \cite{Lemos}. It follows that the constraint matrix is
\begin{equation}
\label{matrix-constraints-counterexample}
\boldsymbol{\mathsf{A}} = 
                   \left( \begin{array}{ccc}
                    1 & 0   &  0   \\                         
										0 & 1   &  0                  
                  \end{array}
                  \right)  
\end{equation}
and it is clear that $\boldsymbol{\mathsf{A}}$ has rank $2$. The matrix $\boldsymbol{\mathsf{B}}$ constructed from $\boldsymbol{\mathsf{A}}$ and $\boldsymbol{\mathsf{W}}$ according to equation  (\ref{matriz-B}) is
\begin{equation}
\label{matrix-B-counterexample}
\boldsymbol{\mathsf{B}} = 
                   \left( \begin{array}{ccccc}
                    0 & 0   &  1 & 0 & 0  \\                         
										0 & 0   &  0 & 1 & 0   \\  
										1 & 0   &  1 & 1 & 0   \\  
										0 & 1   &  1 & 1 & 1   \\  
										0 & 0   &  0 & 1 & 0    								
                  \end{array}
                  \right)  
\end{equation}
It is easy to verify that  $\det \boldsymbol{\mathsf{B}} =0$. However, the Lagrange multipliers can be expressed in terms of  coordinates and velocities alone. In order to check this, we write the equations of motion (\ref{equacoes-Lagrange-com-multiplicadores-de-Lagrange}) for the system:
\begin{eqnarray}
\label{eq-motion-counterexample1}
{\ddot q}_1 + {\ddot q}_2 +  q_1  &  = & \lambda_1 , \\
\label{eq-motion-counterexample2}
{\ddot q}_1 + {\ddot q}_2 + {\ddot q}_3 +  q_2 &  = & \lambda_2\,  \\
\label{eq-motion-counterexample3}
{\ddot q}_2 +  q_3 &  = & 0 .
\end{eqnarray} 
From the constraint equations (\ref{constraints-counterexample}) we find that $ {\ddot q}_1 + {\ddot q}_2 =0$, which yields $\lambda_1 = q_1$. We find, then, that $\lambda_2 ={\ddot q}_3 + q_2$. Differentiating the second of constraint equations (\ref{constraints-counterexample}) with respect to time we have ${\ddot q}_2= {\dot q}_3$, as consequence of which equation (\ref{eq-motion-counterexample3}) becomes  ${\dot q}_3 + q_3=0$.  Therefore, ${\ddot q}_3 = -{\dot q}_3$ implying $\lambda_2 = -{\dot q}_3 + q_2$.  Both Lagrange multipliers
have been expressed solely  in terms of coordinates and velocities. This  shows that   if (\ref{consistency-requirement}) fails it still may be possible to express the multipliers in terms of coordinates and velocities only. However, the failure of (\ref{consistency-requirement}) brings about a different kind of predicament pointed out in Section \ref{Consistency-Requirement}. We have just noticed that the coordinate $q_3$ obeys ${\dot q}_3 + q_3=0$, whose general solution is $q_3(t)=Ce^{-t}$. Substituting this into the constraint equations (\ref{constraints-counterexample}) and performing trivial integrations we arrive at the general solution to the equations of motion:
\begin{eqnarray}
\label{eq-motion-counterexample1-solution}
q_1(t) &  = & A + Ce^{-t}, \\
\label{eq-motion-counterexample2-solution}
q_2(t) &  = & B - Ce^{-t},  \\
\label{eq-motion-counterexample3-solution}
q_3(t) &  = & Ce^{-t}.
\end{eqnarray} 
Note that the general solution to the equations of motion contains only three arbitrary constants, $A, B,\, C$,  but there are four initial conditions to be satisfied, since, due to the constraints  (\ref{constraints-counterexample}), the initial velocities ${\dot q}_1(0)$ and ${\dot q}_2(0)$ are given by $-q_3(0)$ and $q_3(0)$, respectively. In particular, initial conditions such that $q_3(0)=0$
and ${\dot q_3}(0) \neq 0$ cannot be satisfied.

In this example, the appearence of the additional constraint ${\dot q}_3 + q_3=0$ illustrates the general discussion in Section \ref{Consistency-Requirement} about what can happen when  $\det \boldsymbol{\mathsf{B}} =0$.  
With conceivable but probably rare exceptions,  the failure of  condition (\ref{consistency-requirement}) implies that the method of Lagrange multipliers is not fully consistent.

The possibility remains  that the constraints  give rise to a nonsingular matrix $\boldsymbol{\mathsf{B}}$ in spite of the non-positivity of $\boldsymbol{\mathsf{W}}$.    The next example illustrates exactly such  a case.

{\bf Example 2.} Consider the system described by the  Lagrangian 
\begin{equation}
\label{lagrangian-counterexample2}
L = \frac{{\dot q}_1^2}{2} + \frac{{\dot q}_2^2}{2} - \frac{{\dot q}_3^2}{2} - \frac{1}{2} (q_1^2 + q_2^2  + q_3^2 )
\end{equation} 
now under the single nonholonomic constraint
\begin{equation} 
\label{constraints-counterexample2}
{\dot q}_1 + q_3{\dot q}_2  = 0.
\end{equation} 
The Hessian matrix is $\boldsymbol{\mathsf{W}} = \mbox{diag}\, (1,1,-1)$. The constraint matrix is
\begin{equation}
\label{matrix-constraints-counterexample2}
\boldsymbol{\mathsf{A}} = 
                   \left( \begin{array}{ccc}
                    1 & q_3  &  0         									                  
                  \end{array}
                  \right),
\end{equation}
and its rank is $1$ everywhere. Now we have
\begin{equation}
\label{matrix-B-counterexample2}
\boldsymbol{\mathsf{B}} = 
                   \left( \begin{array}{cccc}
                    0 &  1 & q_3 & 0  \\                         
										1 &  1 & 0 & 0   \\  
										q_3 &  0 & 1 & 0   \\  
										0 &  0 & 0 & -1      								
                  \end{array}
                  \right)  
\end{equation}
with $\det \boldsymbol{\mathsf{B}} =1+q_3^2$, which is everywhere nonvanishing. Therefore,  the Lagrange multiplier can be determined exclusively in terms  of the coordinates and velocities. Let us show how to do this explicitly. In the present case the equations of motion (\ref{equacoes-Lagrange-com-multiplicadores-de-Lagrange})  take the following form:
\begin{eqnarray}
\label{eq-motion-counterexample21}
{\ddot q}_1 +  q_1  &  = & \lambda , \\
\label{eq-motion-counterexample22}
{\ddot q}_2  +  q_2 &  = & \lambda q_3 ,  \\
\label{eq-motion-counterexample23}
-{\ddot q}_3 +  q_3 &  = & 0 .
\end{eqnarray} 
From the constraint equation  (\ref{constraints-counterexample2}) it follows that 
\begin{equation} 
\label{constraints-counterexample2-derivative}
{\ddot q}_1 = - q_3{\ddot q}_2 - {\dot q}_2 {\dot q}_3.
\end{equation} 
Now insert this into Eq.(\ref{eq-motion-counterexample21}) and sum the resulting
equation with Eq.(\ref{eq-motion-counterexample22}) previously multiplied by $q_3$ to obtain 
\begin{equation} 
\label{Lagrange-multiplier-counterexample2}
\lambda = \frac{q_1 + q_2q_3 -{\dot q}_2 {\dot q}_3}{1+q_3^2}.
\end{equation} 

\bigskip

From our main result in Section \ref{consistency-proof} and these two examples one infers that  the positivity of the Hessian matrix $\boldsymbol{\mathsf{W}}$ together with the full rank of the constraint matrix  $\boldsymbol{\mathsf{A}}$  is a sufficient condition that guarantees the consistency of the method of Lagrange multipliers in classical mechanics, but   it is not a necessary condition because the method can be  consistent even though $\boldsymbol{\mathsf{W}}$ is not positive.

The previous examples are obviously contrived. Is there any physical theory in which there appears a Lagrangian with  a nonpositive Hessian matrix? In fact, there is. In Weber's electrodynamics, which has some adherents today, the generalized potential (in Gaussian units) that yields the force on a moving point charge $e$ exerted by a point charge $e^{\prime}$ fixed at the origin is  \cite{Whittaker}
\begin{equation} 
\label{Weber-potential}
U(r, {\dot r}) = \frac{ee^{\prime}}{r} \Bigl( 1+ \frac{{\dot r}^2}{2c^2}\Bigr).
\end{equation} 
The Lagrangian in spherical coordinates is 
\begin{eqnarray}
\label{Lagrangiana-particula-potencialcentral}
L & = & \frac{m}{2}
( {\dot r}^2 + r^2 {\dot \theta}^2 + r^2\,{\dot \phi}^2 \sin^2\theta\,
 ) - U(r, {\dot r}) \nonumber \\
& = & \frac{m}{2}
\bigg( 1 -  \frac{ee^{\prime}}{mc^2r}\bigg) {\dot r}^2 + \frac{m}{2}r^2 {\dot \theta}^2 + \frac{m}{2} r^2\,{\dot \phi}^2 \sin^2\theta\,
  - \frac{ee^{\prime}}{r}.
\end{eqnarray}
The Hessian matrix associated with this Lagrangian is not positive if 
\begin{equation} 
\label{r-nonpositive}
 r \leq \frac{ee^{\prime}}{mc^2}.
\end{equation} 
For these values of $r$ this theory  gives rise to physical inconsistencies \cite{Whittaker}.

As we have just seen, if $\boldsymbol{\mathsf{W}}$ is not positive the method of Lagrange multipliers can be consistent, in the sense that the constraint forces are uniquely determined as functions solely of $q, {\dot q},t$.    
Nonetheless, a Lagrangian with a nonpositive Hessian matrix must  be shunned,
for it is almost certainly bound to give rise to some other kind of physical inconsistency

\section{Conclusions} 

The interplay between mathematics and physics has long and often been the cause of wonder and pondering for prominent mathematicians and theoretical physicists.  Wigner \cite{Wigner} justifiably marveled at the fact that  ``the mathematical formulation of the physicist's often crude experience leads in an uncanny
number of cases to an amazingly accurate description of a large class of phenomena.'' Dirac \cite{Dirac} ascribed the remarkable success of the method of mathematical reasoning in physics to some mathematical quality in nature. To him the principle of mathematical beauty plays a central role in the setting up of physical theories.  V. I. Arnold \cite{Arnold} viewed mathematics as a natural science, and famously stated that ``mathematics is the part of physics where experiments are cheap.''

It is not  always appreciated  that  scrutinizing the mathematical foundations of a physical theory may help identify  conditions  that must be imposed to add physical content to the mathematical setting. 
In order to put a physical theory on a rigorous mathematical basis it is not enough to explicitly state conditions under which every problem admits a unique solution. Mathematical existence does not by itself guarantee that the solution is physically reasonable. Existence and uniqueness of a {\it physical} solution may require  conditions additional to those that ensure mathematical existence and uniqueness.  Rigorous mathematical formulations are helpful to clarify the physics, and the physics assists in suggesting the most suitable  mathematical framework.

We have presented a sufficient  condition, namely equation (\ref{consistency-requirement}),  that the Lagrangian and the constraints must satisfy in order to guarantee that we can solve for the Lagrange multipliers and obtain the physical constraint forces. We have also cogently argued that the failure of (\ref{consistency-requirement}) is nearly certain to preclude the method of Lagrange multipliers from being fully consistent. Moreover, through the examination of an important physical topic we hope to have shown that, even though our physical intuition is fundamental for the treatment of different classes of problems, there are cases in which a careful mathematical analysis is needed to show the consistency, or lack thereof, of our own intuition.

\begin{acknowledgments}
We would like to thank the anonymous referees and the editor Todd Springer for their careful reading and constructive criticism, which helped improve our paper considerably.
\end{acknowledgments}

\begin{appendix}
\section*{Appendix: Vakonomic mechanics}

Vakonomic mechanics \cite{Ray,Goedecke,Kozlov,Borisov} is an approach to the dynamics of systems subject to  velocity-dependent constraints of the general form (\ref{non-holon-more-general}) that differs from the d'Alembert-Lagrange theory described in Section \ref{Lagrangemultipliermethod}. Its inspiration comes from  the mathematical problem of Lagrange in the calculus of variations \cite{Bliss}.
The equations of motion of vakonomic mechanics arise from the variational principle
 \begin{equation}
\label{vakonomic-variational-principle}
\delta \int_{t_1}^{t_2}\Bigl[ L(q, {\dot q},t) + \sum_{l=1}^p \lambda_l(t) g_l(q, {\dot q},t) \Bigr] dt = 0  ,
\end{equation} 
in which the $q$s and $\lambda$s are varied independently, with the standard boundary conditions $\delta q_k(t_1) =  \delta q_k(t_2) = 0$. Variation of the Lagrange multipliers  leads to the constraint equations (\ref{non-holon-more-general}), whereas variation of the coordinates yields the equations of motion 
\begin{equation}
\label{vakonomic-equations-of-motion}
\frac{d}{dt}
\Bigl(\frac{\partial L}{\partial  {\dot q_k}} \Bigr) -
\frac{\partial L}{\partial q_k} =
\sum_{l=1}^p\,{\lambda}_l \bigg[ \frac{\partial g_l}{\partial q_k} - \frac{d}{dt}
\Bigl(\frac{\partial g_l}{\partial  {\dot q_k}} \Bigr) \bigg] - \sum_{l=1}^p {\dot \lambda_l}\frac{\partial g_l}{\partial  {\dot q_k}}
\, , \,\,\,\,\,\,\,\,\,\,
k=1,\ldots ,n\, .
\end{equation}
For holonomic constraints
\begin{equation}
\label{holonomic-constraints}
f_l(q,t) =0, \,\,\,\,\,\,\,\,\,\, l=1,\ldots ,p,
\end{equation}
we have 
\begin{equation}
\label{g-holonomic-constraints}
g_l(q, {\dot q},t) = \frac{d}{dt} f_l(q,t) = \sum_{j=1}^n \frac{\partial f_l}{\partial q_j}{\dot q}_j + \frac{\partial f_l}{\partial t},
\end{equation}
and it is easy to check  that
\begin{equation}
\label{g-variational-derivative-zero}
\frac{\partial g_l}{\partial q_k} - \frac{d}{dt}
\Bigl(\frac{\partial g_l}{\partial  {\dot q_k}}\Bigr) =0.
\end{equation}
This reduces the vakonomic equations of motion  (\ref{vakonomic-equations-of-motion}) to
\begin{equation}
\label{vakonomic-equations-of-motion-holonomic}
\frac{d}{dt}
\Bigl(\frac{\partial L}{\partial  {\dot q_k}} \Bigr) -
\frac{\partial L}{\partial q_k} =
 - \sum_{l=1}^p {\dot \lambda_l}\frac{\partial g_l}{\partial  {\dot q_k}}
\, , \,\,\,\,\,\,\,\,\,\,
k=1,\ldots ,n\, .
\end{equation}
Since only ${\dot \lambda_l}$ appear in these equations but not the Lagrange multipliers $ \lambda_l$ themselves, the mere renaming  ${\dot \lambda_l} \to -\lambda_l$ shows that the equations of motion of  vakonomic mechanics coincide with (\ref{equacoes-Lagrange-com-multiplicadores-de-Lagrange-general}) if the constraints are holonomic.

However, if  the constraints are nonholonomic, equations   (\ref{vakonomic-equations-of-motion}) and (\ref{equacoes-Lagrange-com-multiplicadores-de-Lagrange-general}) 
are completely different. Although mathematical conditions are known \cite{Fernandez} under which equations (\ref{vakonomic-equations-of-motion}) and (\ref{equacoes-Lagrange-com-multiplicadores-de-Lagrange-general}) can be regarded as equivalent, the presence  of both $\lambda_l $ and ${\dot \lambda_l} $ in the equations of motion of   vakonomic mechanics means that in order to determine the motion uniquely one has to specify $\lambda_l(0)$, the values  of the  Lagrange multipliers at the initial instant $t=0$. This entails, given that the Lagrange multipliers are directly connected to the constraint forces, that in general, though not always, the initial accelerations would have to be prescribed together with the initial positions and velocities, which is not physically sound. Therefore, the physical significance of vakonomic mechanics, if any, remains obscure \cite{Flannery2,Flannery3}. Moreover, experimental studies \cite{Lewis,Kai} favor (\ref{equacoes-Lagrange-com-multiplicadores-de-Lagrange-general}) as the correct equations of motion for nonholonomic systems.

This brief critical exposition of vakonomic mechanics is justified because there is an ongoing controversy concerning its physical validity. Moreover, to add to the confusion, equations (\ref{vakonomic-equations-of-motion}) were sanctioned in the third edition of Goldstein's highly influential textbook \cite{Goldstein} but retracted without explanation by the new co-authors almost as soon as it came to light \cite{Goldstein2}.

\end{appendix}


\begin{thebibliography}{99}


\bibitem{Lemos}  N. A. Lemos, {\it Analytical Mechanics} (Cambridge University Press, Cambridge, 2018).



\bibitem{English} L. Q. English and A. Mareno, ``Trajectories of rolling marbles on various funnels,'' Am. J. Phys. {\bf 80}, 996-1000 (2012).

\bibitem{White} G. D. White, ``On trajectories of rolling marbles in cones and other funnels,'' Am. J. Phys. {\bf 81}, 890-898 (2013).

\bibitem{Neimark} Ju. I. Ne\u{\i}mark and N. A. Fufaev, {\it Dynamics of Nonholonomic Systems} (American Mathematical Society, Providence, RI,  1972).

\bibitem{Gatland} I. R. Gatland, ``Nonholonomic constraints: A test case,'' Am. J. Phys. {\bf 72}, 941-942 (2004).

\bibitem{Janova} J. Janov\'a and J. Musilov\'a, ``The streetboard rider: an appealing
problem in non-holonomic mechanics,'' Eur. J. Phys. {\bf 31}, 333–345 (2010).

\bibitem{Zangwill} Thus we are  excluding self-interaction forces that arise in electromagnetism, such as the Abraham-Lorentz force. See, for example, A. Zangwill, {\it Modern Electrodynamics} (Cambridge University Press, Cambridge, 2013).

\bibitem{Rund} H. Rund, {\it The Hamilton-Jacobi Theory in the Calculus of Variations} (Van Nostrand, London, 1966).

\bibitem{Saletan} E. J. Saletan and A. H. Cromer, ``A variational principle for nonholonomic systems,'' Am. J. Phys. {\bf 38}, 892-897 (1970).



\bibitem{Flannery1} M. R. Flannery, ``The enigma of nonholonomic constraints,'' Am. J. Phys. {\bf 73}, 265-272 (2005).

\bibitem{Flannery2} M. R. Flannery, ``d'Alembert–Lagrange analytical dynamics for
nonholonomic systems,'' J. Math. Phys. {\bf 52}, 032705 (2011).

\bibitem{Flannery3} M. R. Flannery, ``The elusive d’Alembert-Lagrange dynamics of nonholonomic systems,'' Am. J. Phys. {\bf 79}, 932-944 (2011).

\bibitem{Bliss} G. A. Bliss, ``The problem of Lagrange in the calculus of variations,'' Amer. J. Math. {\bf 52}, 673-744 (1930).

\bibitem{Ray} J. R. Ray, “Nonholonomic Constraints,” Am. J. Phys. {\bf 34}, 406–408
(1966); Erratum  {\bf 34}, 1202–1203 (1966).

\bibitem{Goedecke} G. H. Goedecke, ``Undetermined multiplier treatments of the Lagrange
problem,'' Am. J. Phys. {\bf 34}, 571–574 (1966).

\bibitem{Kozlov} V. V. Kozlov, ``Realization of nonintegrable constraints in classical mechanics,'' Sov. Phys. Dokl. {\bf 28}, 735-737
(1983).

\bibitem{Borisov} A. V. Borisov, I. S. Mamaev and I. A. Bizyaev, ``Dynamical systems with non-integrable constraints,
vakonomic mechanics, sub-Riemannian
geometry, and non-holonomic mechanics,'' Russ. Math. Surv. {\bf 72}, 783-840 (2017).



\bibitem{Marsden}  J. E. Marsden and T. S. Ratiu, {\it Introduction to  Mechanics and Symmetry}, 2nd ed. (Springer, New York, 1999).


\bibitem{Whittaker}  E. T. Whittaker,  {\it A History of the Theories of Aether and Electricity, Vol. I:  The Classical Theories} (Thomas Nelson and Sons,  London, 1951), Chap. VII.

\bibitem{Wigner} E. Wigner,  ``The unreasonable effectiveness of mathematics in the natural sciences,'' Commun. Pure Appl.  Math. {\bf 13}, 1-14 (1960).

\bibitem{Dirac} P. A. M. Dirac, ``The Relation between Mathematics and Physics,'' Proc. Royal Soc. (Edinburgh) {\bf  59}, 122-129 (1938-39).

\bibitem{Arnold} V. I. Arnold, ``On teaching mathematics,'' address at the Palais de D\'ecouverte in Paris, March 7, 1997. Russian Math. Surveys {\bf 53}, 229-236 (1998).
https://www.uni-muenster.de/Physik.TP/$\%$7Emunsteg/arnold.html.

\bibitem{Fernandez} O. E. Fernandez and A. M. Bloch, ``Equivalence of the dynamics of nonholonomic and
variational nonholonomic systems for certain initial
data,'' J. Phys. A: Math. Theor. {\bf 41},   344005 (2008).


\bibitem{Lewis} A. D. Lewis and R. M. Murray, ``Variational principles for constrained systems: Theory and experiment,'' Int. J. Non-Linear Mechanics {\bf  30}, 793-815 (1995).


\bibitem{Kai} T. Kai, ``Experimental comparison between nonholonomic and vakonomic mechanics in nonlinear constrained systems,'' Nonlinear Theory and Its Applications 
 {\bf 4}, 482-499 (2013).


\bibitem{Goldstein}  H. Goldstein, C. P. Poole, Jr. and J. L. Safko, {\it Classical Mechanics}, 3rd ed. (Addison Wesley, San Francisco, 2001).

\bibitem{Goldstein2}  http://astro.physics.sc.edu/Goldstein/1-2-3To6.html. At this website,  devoted to errata, corrections and comments on the third edition, co-author J. L. Safko  states that Section 2.4  should be revised and cites, among others, reference \cite{Saletan} above.

\end{thebibliography}
\end{document}